\documentclass[american,12pt]{article}
\usepackage{pdfpages}
\usepackage{setspace}
\usepackage{tikz}
\usepackage{epsfig}
\usepackage{amsmath}
\usepackage{amsfonts}
\usepackage{pst-plot}
\usepackage{hyperref}
\usepackage{amssymb}
\usepackage{pgffor}
\usepackage[margin=1.25in]{geometry}
\usepackage{babel}
\usepackage{subfig}
\usepackage{tikz}
\usepackage{booktabs,siunitx}
\usepackage{accents}
\usepackage{mathrsfs}
\usepackage{amsfonts}
\usepackage{amssymb}
\usepackage{pgffor}
\usepackage{eufrak}
\usepackage[utf8]{inputenc}
\usepackage{enumitem}
\usepackage{amsmath,fouriernc}
\usepackage{comment}

\newcommand*{\field}[1]{\mathbb{#1}}

\usetikzlibrary {positioning}

\definecolor {processblue}{cmyk}{0.96,0,0,0}
\usetikzlibrary{arrows}
\usepackage[authoryear]{natbib}
\input{glyphtounicode}
\DeclareMathAlphabet{\mathpzc}{OT1}{pzc}{m}{it}

\makeatletter
\renewcommand{\section}{\@startsection{section}{1}{0mm}{-1.5\baselineskip}{0.8\baselineskip}{\normalfont\large\centering}}
\renewcommand{\subsection}{\@startsection{subsection}{2}{0mm}{-0.1\baselineskip}{0.5\baselineskip}{\normalfont\bf\flushleft}}
\renewcommand{\@seccntformat}[1]{\csname the#1\endcsname \hspace{+0mm}\large{.}\hspace{+1.9mm}}
\renewcommand{\@seccntformat}[2]{\csname the#1\endcsname \hspace{+0mm}\large{.}\hspace{+1.9mm}}
\makeatother

\newtheorem{theorem}{Theorem}

\newtheorem{definition}{Definition}

\newtheorem{proposition}{Proposition}
\newtheorem{remark}{Remark}

\renewcommand{\theequation}{\arabic{equation}}

\newlength{\extraspace}
\newlength{\extraspaces}
\newcounter{dummy}

\newcommand{\baa}{
\addtocounter{equation}{1} \setcounter{dummy}{\value{equation}}
\setcounter{equation}{0}
\renewcommand{\theequation}{\arabic{dummy}\alph{equation}}
\begin{eqnarray}
\addtolength{\abovedisplayskip}{\extraspaces}
\addtolength{\belowdisplayskip}{\extraspaces}
\addtolength{\abovedisplayshortskip}{\extraspace}
\addtolength{\belowdisplayshortskip}{\extraspace}}

\newcommand{\eaa}{
\end{eqnarray}
\setcounter{equation}{\value{dummy}}
\renewcommand{\theequation}{\arabic{equation}}}

\newcommand{\be}{\begin{equation}
\addtolength{\abovedisplayskip}{\extraspaces}
\addtolength{\belowdisplayskip}{\extraspaces}
\addtolength{\abovedisplayshortskip}{\extraspace}
\addtolength{\belowdisplayshortskip}{\extraspace}}
\newcommand{\ee}{\end{equation}}

\newcommand{\ba}{\begin{eqnarray}
\addtolength{\abovedisplayskip}{\extraspaces}
\addtolength{\belowdisplayskip}{\extraspaces}
\addtolength{\abovedisplayshortskip}{\extraspace}
\addtolength{\belowdisplayshortskip}{\extraspace}}
\newcommand{\ea}{\end{eqnarray}}

\newcommand{\bd}{\begin{displaymath}
\addtolength{\abovedisplayskip}{\extraspaces}
\addtolength{\belowdisplayskip}{\extraspaces}
\addtolength{\abovedisplayshortskip}{\extraspace}
\addtolength{\belowdisplayshortskip}{\extraspace}}
\newcommand{\ed}{\end{displaymath}}

\def\inbar{\,\vrule height1.5ex width.4pt depth0pt}
\font\rms=cmr12 at 12pt
\def\ce{\relax\ifmmode\mathchoice
{\hbox{$\inbar\kern-.3em{\rm C}$}} {\hbox{$\inbar\kern-.3em{\rm
C}$}} {\lower.9pt\hbox{\rms $\inbar\kern-.3em{\rm C}$}}
{\lower1.2pt\hbox{\rms $\inbar\kern-.3em{\rm C}$}}
\else{$\inbar\kern-.3em{\rm C}$}\fi}
\font\cmss=cmss12 \font\cmsss=cmss12 at 12pt
\def\ze{\relax\ifmmode\mathchoice
{\hbox{\cmss Z\kern-.4em Z}}{\hbox{\cmss Z\kern-.4em Z}}
{\lower.9pt\hbox{\cmsss Z\kern-.4em Z}} {\lower1.2pt\hbox{\cmsss
Z\kern-.4em Z}}\else{\cmss Z\kern-.4em Z}\fi}

\newcommand{\refsection}[1]{
\vspace{1mm} \pagebreak[3] \addtocounter{section}{1}
\begin{center}
{\large #1}
\end{center}
\nopagebreak
\medskip
\nopagebreak}

\def\thebibliography#1{\refsection{\bf References}
\vspace*{-8mm}\list
 {\relax}{\itemsep=1pt \parsep=0pt
 \usecounter{enumiv}\leftmargin=3em\itemindent=-\leftmargin}%
 \def\newblock{\hskip .11em plus .33em minus .07em}
 \sloppy\clubpenalty4000\widowpenalty4000
 \sfcode`\.=1000\relax}

\newcommand{\q}[1]{``#1''}

\begin{document}

\begin{center}
{\Huge\sc Tropical Analysis}\\[3mm]
{\Large\sc With an Application to Indivisible Goods}\\[15mm]
{\large Nicholas C. Bedard and Jacob K. Goeree}\footnote{Bedard: Wilfrid Laurier University, 75 University Avenue W., Waterloo, Ontario, Canada N2L 3C5, Canada. Goeree: AGORA Center for Market Design and the School of Economics, UNSW Australia Business School, Sydney NSW 2052, Australia; Email: \href{mailto: jacob.goeree@gmail.com}{jacob.goeree@gmail.com}. We would like to thank the Australian Research Council (DP190103888 and DP220102893) for financial support.}\\[5mm]
\today\\[15mm]
{\bf Abstract}
\end{center}
\addtolength{\baselineskip}{-1.2mm}
\vspace*{-3mm}

\noindent We establish the \textit{Subgradient Theorem} for monotone correspondences -- a monotone correspondence is equal to the subdifferential of a potential iff it is conservative, i.e.
its integral along a closed path vanishes irrespective of the selection from the correspondence along the path. We prove two attendant results: the \mbox{\textit{Potential Theorem}}, whereby a conservative monotone correspondence can be integrated up to a potential, and the \textit{Duality Theorem}, whereby the potential has a Fenchel dual whose subdifferential
is another conservative monotone correspondence. We use these results to reinterpret and extend \citeauthor{BaldwinKlemperer2019}'s (\citeyear{BaldwinKlemperer2019}) characterization of demand in economies with indivisible goods. We introduce a simple test for existence of Walrasian equilibrium in quasi-linear economies. Fenchel's Duality Theorem implies this test is met when the aggregate utility is concave, which is not necessarily the case with indivisible goods even if all consumers have concave utilities.

\vfill
\noindent {\bf Keywords}: \textit{Conservative correspondences, subgradient theorem, potential theorem, Fenchel duality, Fenchel's duality theorem, tropical geometry, convex analysis, normally labeled polyhedral subdivisions, subdifferentials, indivisible goods}

\addtocounter{footnote}{-1}
\newpage
\addtolength{\baselineskip}{1.25mm}

\section{Introduction}

Multi-valued vector functions, or correspondences, play an important role in many areas of economics, including consumer and producer choice, general equilibrium, game theory, and mechanism design. For instance, proving existence of equilibrium typically rests on \citeauthor{Kakutani1941}'s (\citeyear{Kakutani1941}) generalization of Brouwer's fixed-point theorem to correspondences. In a variety of optimization problems, \textit{single}-valued solutions satisfy an integrability condition: they are gradients of convex (or concave) functions known as potentials.\footnote{For instance, \citeauthor{Hotelling1932}'s (\citeyear{Hotelling1932}) lemma dictates that the supply function is the gradient of the profit function. \citeauthor{Shephard1970}'s (\citeyear{Shephard1970}) lemma dictates that Hicksian demand is the gradient of the expenditure function and that factor demand is the gradient of the cost function. Finally, with quasi-linear preferences, \citeauthor{Roy1947}'s (\citeyear{Roy1947}) identity dictates that Marshallian demand is minus the gradient of the indirect utility function, see Section 3. In mechanism design, the optimal allocation rule often corresponds to the gradient of a convex potential, see e.g. \cite{RochetChone1998,ManelliVincent2007,GoereeKushnir2023}.} Specifically, a single-valued function is the gradient of a differentiable potential if and only if it is conservative, i.e. its integral along every closed path vanishes. This result is known as the \q{Gradient Theorem.}

When is a correspondence equal to the subdifferential of a potential? \cite{Rockafellar1970} shows that a necessary and sufficient condition is that the correspondence is maximal and cyclically monotone. To relate cyclical monotonicity to the more geometric notion of conservativeness, we build on \citeauthor{Aumann1965}'s (\citeyear{Aumann1965}) work on integrals of correspondences defined on the unit interval. The Aumann integral generally yields a convex set but we show that for a monotone correspondence this set is a singleton. We extend this uniqueness result to closed-path integrals of monotone correspondences in arbitrary dimensions, which allows us to define \textit{conservativeness} -- a monotone correspondence is conservative if and only if its integral along any closed path vanishes -- and establish the \textit{Subgradient Theorem} -- a monotone correspondence is equal to the subdifferential of some potential if and only if it is conservative.

Conversely, the Subgradient Theorem allows us to determine the potential by integrating the conservative monotone correspondence -- the \textit{Potential Theorem}. The usefulness of the Potential Theorem is that the Fenchel dual of a convex (concave) potential is a concave (convex) potential and its subdifferential defines a maximal conservative monotone correspondence inverse to the original correspondence -- the \textit{Duality Theorem}. This theorem allows us to solve problems in one domain that seem intractable in the dual domain.

We use our duality result in an application that is based on an inspiring recent contribution by \cite{BaldwinKlemperer2019} who use insights from tropical geometry to study consumer choice when goods are indivisible and preferences are quasi-linear. They show that the graph of Marshallian demand in price space forms a polyhedral complex, the \q{price complex,} and that its facets, on which demand is multi-valued, form a tropical curve.\footnote{A polyhedral complex is a collection of polygons that form a partition of the space. The tropical graph is where the polygons meet.} Conversely, they show that a price complex arises from maximization of a concave utility if and only if the tropical curve formed by its weighted facets is \q{balanced} \citep{Mikhalkin2004}. Finally, they show that the price complex is dual to a polyhedral complex in quantity space, the \q{demand complex,} but note that its nature is more \q{abstract} in the sense that it is associated to a class of utility functions rather than a specific valuation.

We show that the geometric duality outlined by \cite{BaldwinKlemperer2019} is a reflection of the usual Fenchel duality between demand and inverse demand and that balancedness of their price and demand complexes is a reflection of conservativeness of demand and inverse demand. Specifically, we show that the price complex is equivalent to the subdifferential of indirect utility -- the former is the vertical projection of the latter onto its domain -- and that \citeauthor{Mikhalkin2004}'s (\citeyear{Mikhalkin2004}) balancing condition is a tropical version of conservativeness. Since the inverse of a subdifferential correspondence is itself a subdifferential correspondence, its vertical projection defines another polyhedral complex dual to the original. This dual complex is equivalent to the subdifferential of utility and generalizes \citeauthor{BaldwinKlemperer2019}'s (\citeyear{BaldwinKlemperer2019}) demand complex in that it is not \q{abstract,} but contains the same information as the price complex. We relate our findings to classical results in consumer demand theory, e.g. in price space, demand follows from \citeauthor{Roy1947}'s (\citeyear{Roy1947}) identity and in quantity space inverse demand follows from the usual premise that marginal utilities equal prices.

Finally, we exploit our duality results to provide a simple test whether Walrasian equilibrium exists in economies with quasi-linear preferences. A Walrasian equilibrium exists if and only if the minimum of aggregate indirect utility over prices is equal to the maximum of its dual, aggregate utility, over allocations. When aggregate utility is concave, existence does not require any fixed-point arguments but follows from Fenchel's Duality Theorem. However, in economies with indivisible goods, the aggregate utility function is not necessarily concave even if all individual utility functions are. As a result, Walrasian equilibrium may fail to exist and our duality test provides a simple way to check for existence.

Section \ref{sec:gen-theory} defines conservative monotone correspondences and derives the Subgradient, Potential, and Duality Theorems. Section 3 reexamines demand with indivisible goods. Section 4 discusses equilibrium in economies with quasi-linear preferences and indivisible goods. Section 5 concludes. The Appendices contain the proofs.

\section{Conservative Monotone Correspondences}\label{sec:gen-theory}

In this section we show that a monotone correspondence is equal to the subdifferential of some potential if and only if it is conservative. We also characterize this potential and establish properties of its dual. These are novel result in the mathematics literature and in Section \ref{sec:indv-goods} we apply them to study important economic environments where differentiability cannot be assumed. To emphasize the economic application of these tools, we'll use the following notation:
\begin{itemize}\addtolength{\itemsep}{-2mm}
\vspace*{-1mm}
\item[--] $V(\mathbf{p})$ is a convex function with subdifferential $\mathbf{Q}(\mathbf{p})=-\partial V(\mathbf{p})$.
\item[--] $U(\mathbf{q})$ is a concave function with subdifferential $\mathbf{P}(\mathbf{q})=\partial U(\mathbf{q})$.
\end{itemize}
The standard economic interpretation is that $V(\mathbf{p})$ is the indirect utility function, which is convex, and $\mathbf{Q}(\mathbf{p})$ is the (Marshallian) demand correspondence (via Roy's lemma). Likewise, $U(\mathbf{q})$ is the utility function, which is concave, and $\mathbf{P}(\mathbf{q})$ is the inverse demand correspondence.

More generally, let $\mathbf{Q}:C\rightarrow\field{R}^n$ be a correspondence defined on some convex domain $C\subseteq\field{R}^n$. Recall that $\mathbf{Q}$ is \textit{monotone} if and only if
\begin{displaymath}
  (\mathbf{q}_2-\mathbf{q}_1)\cdot(\mathbf{p}_2-\mathbf{p}_1)\,\leq\,0
\end{displaymath}
for all $\mathbf{p}_1,\mathbf{p}_2\in C$, $\mathbf{q}_1\in\mathbf{Q}(\mathbf{p}_1)$, $\mathbf{q}_2\in\mathbf{Q}(\mathbf{p}_2)$.\footnote{We define a monotone correspondence as decreasing to accord with our economic interpretation. In the literature on convex analysis the inequality is typically reversed, defining an increasing correspondence. Also, some authors use the terminology superdifferential for concave functions. We follow \cite{Rockafellar1970} who uses subdifferential for both convex and concave functions.}  A correspondence is maximal if its graph is not properly contained in the graph of another monotone correspondence.\footnote{If $Q$ is not maximal we can replace it with its maximal extension, $\overline{Q}$, see footnote \ref{fn:max}. Intuitively, in the univariate case, $\overline{Q}$ is such that its graph can be obtained by drawing a curve through the graph of $Q$ without lifting one's pen from the paper.}

\begin{definition}
A maximal monotone correspondence, $\mathbf{Q}$, is conservative if and only if, for any closed path $\Gamma\subset C$,
\begin{equation}\label{def:conservative}
  \oint_{\Gamma}\,\mathbf{Q}(\mathbf{p})\cdot d\mathbf{p}\,=\,0
\end{equation}
\end{definition}

\medskip

\noindent We show in Appendix \ref{App:A} that the path integral in \eqref{def:conservative} is unique, i.e. independent of the selection from $\mathbf{Q}$ used to compute it, and therefore \eqref{def:conservative} is well defined.
\begin{theorem}[Subgradient Theorem]\label{th:subgrad}
If $\,\,\,V:C\rightarrow\field{R}^n$ is a non-increasing and convex function then $\mathbf{Q}=-\partial V$ is a maximal conservative monotone correspondence on $C$. Conversely, if $\,\mathbf{Q}$ is a maximal monotone conservative correspondence on $C$ then $\mathbf{Q}=-\partial V$ for some non-increasing and convex function $V:C\rightarrow\field{R}^n$.
\end{theorem}

\begin{remark}{\em
As convex functions are differentiable almost everywhere, the gradient $\nabla V$ is the unique section from the subdifferential $\partial V$ almost everywhere. But one cannot simply use the standard gradient theorem to establish conservativeness because the path $\Gamma$ might be partly, or completely, contained in a lower-dimensional set of non-differentiability. For instance, if $V(p_1,p_2)=-\max(p_1^2+p_2^2,1)$ and $\Gamma$ is the unit circle, then the gradient does not exist anywhere along the path. However, the difference between any two selections from $\partial V(p_1,p_2)=-\alpha (p_1,p_2)$ for $0\leq\alpha\leq 2$ is a vector normal to the path. So while $\partial V$ is multi-valued, the inner product of $\partial V$ with the path's tangent is 0, i.e. \textit{single valued}, along the entire path. Hence, different selections for the subgradients yield the same result for the path integral in \eqref{def:conservative}. See the proof of Theorem \ref{th:subgrad} for further details.}$\hfill\blacksquare$
\end{remark}
We can be more specific about the \q{some convex function $V$} in the converse part of Theorem \ref{th:subgrad}. Pick some $\mathbf{p}_0\in C$ and some finite value for $V(\mathbf{p}_0)$. Let $\Gamma(\mathbf{p}_0,\mathbf{p})$ denote a path from $\mathbf{p}_0$ to $\mathbf{p}$. If $\mathbf{Q}$ is conservative,
\begin{equation}\label{potential}
 V(\mathbf{p})\,=\,V(\mathbf{p}_0)-\int_{\Gamma(\mathbf{p_0},\mathbf{p})}\,\mathbf{Q}(\mathbf{p})\cdot d\mathbf{p}
\end{equation}
is independent of $\Gamma(\mathbf{p}_0,\mathbf{p})$. This function is known as the \textit{potential} for $\mathbf{Q}$.
\begin{theorem}[Potential Theorem]\label{th:potential}
Let $\mathbf{Q}$ be a maximal conservative monotone correspondence and let the potential $V$ be defined as in \eqref{potential} then $\mathbf{Q}=-\partial V$.
\end{theorem}

\medskip

\noindent An important property of subdifferential mappings is that they can be inverted in the sense of multi-valued mappings, see \citeauthor{Rockafellar1970} (\citeyear[Cor. 23.5.1]{Rockafellar1970}). If $\mathbf{Q}=-\partial V$ then there exists another maximal monotone correspondence, $\mathbf{P}= \partial U$, such that $\mathbf{p}\in\mathbf{P}(\mathbf{q})$ if and only if $\mathbf{q}\in\mathbf{Q}(\mathbf{p})$. Here $U$ is the Fenchel dual of $V$:\footnote{Typically, the Fenchel dual of a convex function $f(\mathbf{p})$ is defined to be another convex function $f^*(\mathbf{q})=\max_\mathbf{p}\,\langle\mathbf{p}|\mathbf{q}\rangle-f(\mathbf{p})$, see \cite{Rockafellar1970}. The definition of Fenchel duality used here maintains the economics convention that duality is between a concave utility and a convex indirect utility.}
\begin{equation}\label{FenchelDual}
  U(\mathbf{q})\,=\,\min_{\mathbf{p}}\,V(\mathbf{p})+\langle\mathbf{p}\,|\,\mathbf{q}\rangle
\end{equation}
Since $\mathbf{P}$ is a subdifferential mapping it is a maximal conservative monotone correspondence by Theorem \ref{th:subgrad}.
\begin{theorem}[Duality]\label{th:duality}
Let $\mathbf{Q}$ be a maximal conservative monotone correspondence with potential $V$. Then its inverse $\mathbf{P}= \partial U$ is a maximal conservative monotone correspondence with potential $U$, which is the Fenchel dual of $V$.
\end{theorem}

\medskip

\noindent In physics, potentials are typically assumed to be differentiable everywhere and the duality result is attributed to Legendre ($U$ is called the Legendre transform of $V$). Theorem \ref{th:duality} generalizes the duality result to allow for non-differentiabilities, which naturally arise in economics problems with indivisibilities, e.g. the assignment of a discrete set of goods. In such problems, the potentials are typically finitely generated, or polyhedral.

\section{Economies with Indivisible Goods} \label{sec:indv-goods}

Economies with indivisible goods pose an obvious technical challenge, even with a single consumer. The usual approach of equating marginal rates of substitution to price ratios seems impossible with only a finite set of values for a discrete set of bundles. To deal with this challenge, \cite{BaldwinKlemperer2019} use insights from the mathematics literature on tropical geometry. In this section we show that the duality they exploit, between demand in terms of prices and (inverse) demand in terms of quantities, fits squarely in the domain of convex analysis using our results from Section \ref{sec:gen-theory}.\footnote{The terminology ``tropical analysis'' reflects the overlap between results in tropical geometry and convex analysis.} In particular, the geometric duality they outline is a reflection of the usual Fenchel duality between demand and inverse demand and the balancedness of their price and demand complexes is a reflection of the conservativeness of the demand and inverse demand correspondences.

\begin{figure}[t]
\begin{center}
\vspace*{4mm}
\begin{tikzpicture}[scale=.5]
\fill (0,0) circle (5pt);
\fill (6,0) circle (5pt);
\fill (3,3) circle (5pt);
\fill (0,6) circle (5pt);
\fill (6,6) circle (5pt);
\node[below,scale=0.8] at (0,-1.2) {$U_1=0$};
\node[below,scale=0.8] at (0,-0.2) {$\mathpzc{q}_1=(0,\!0)$};
\node[below,scale=0.8] at (6,-1.2) {$U_2=16$};
\node[below,scale=0.8] at (6,-0.2) {$\mathpzc{q}_2=(2,\!0)$};
\node[below,scale=0.8] at (3,3-1.2) {$U_3=24$};
\node[below,scale=0.8] at (3,3-0.2) {$\mathpzc{q}_3=(1,\!1)$};
\node[below,scale=0.8] at (0,6-1.2) {$U_4=28$};
\node[below,scale=0.8] at (0,6-0.2) {$\mathpzc{q}_4=(0,\!2)$};
\node[below,scale=0.8] at (6,6-1.2) {$U_5=34$};
\node[below,scale=0.8] at (6,6-0.2) {$\mathpzc{q}_5=(2,\!2)$};
\end{tikzpicture}
\end{center}
\vspace*{-2mm}
\caption{Example of a finite bundle set, $Q=\{\mathpzc{q}_k\}_{k=1}^5$, where the value of $\mathpzc{q}_k$ is $u_k$.}\label{fig:discrete}
\vspace*{0mm}
\end{figure}
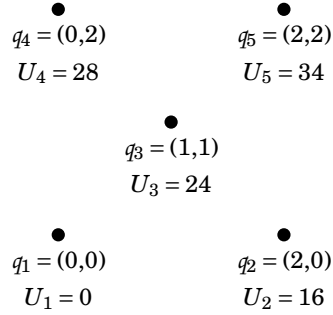

\begin{figure}[t]
\begin{center}
\begin{tikzpicture}[scale=0.8]
\draw[->] (-0.2,0) -- (6,0) node [scale=0.8,right] {$p_1$};
\draw[->] (0,-0.2) -- (0,6) node [scale=0.8,above] {$p_2$};
\draw[line width=1pt] (2,1) -- (4/3,5/3) -- (11/3,13/3) -- (13/3,11/3) -- (2,1);
\draw[line width=1pt] (2,1) -- (2,-1);
\draw[line width=1pt] (4/3,5/3) -- (-1,5/3);
\draw[line width=1pt] (11/3,13/3) -- (11/3,6);
\draw[line width=1pt] (13/3,11/3) -- (6,11/3);

\fill[blue] (-0.5,5/3) circle (8pt);
\fill[blue] (2,-0.5) circle (8pt);
\fill[blue] (11/3,5.5) circle (8pt);
\fill[blue] (5.5,11/3) circle (8pt);
\node[white,scale=0.7] at (-0.5,5/3) {\bf 2};
\node[white,scale=0.7] at (2,-0.5) {\bf 2};
\node[white,scale=0.7] at (11/3,5.5) {\bf 2};
\node[white,scale=0.7] at (5.5,11/3) {\bf 2};

\fill (2,1) circle (4pt);
\fill (4/3,5/3) circle (4pt);
\fill (11/3,13/3) circle (4pt);
\fill (13/3,11/3) circle (4pt);

\node at (0.8,0.7) {\small $(2,\!2)$};
\node at (4,1.5) {\small $(0,\!2)$};
\node at (1.5,4) {\small $(2,\!0)$};
\node at (2.8,2.6) [rotate=45] {\small $(1,\!1)$};
\node at (4.75,4.5) {\small $(0,\!0)$};

\node[scale=0.8] at (1,-0.4) {$0$};
\node[scale=0.8] at (3,-0.4) {$6$};
\node[scale=0.8] at (5,-0.4) {$12$};
\node[scale=0.8] at (-0.4,0) {$4$};
\node[scale=0.8] at (-0.4,8/3) {$12$};
\node[scale=0.8] at (-0.4,16/3) {$20$};
\end{tikzpicture}
\hspace{2cm}
\begin{tikzpicture}[scale=0.8]
\draw[line width=1pt] (0,0) -- (6,0) -- (6,6) -- (0,6) -- (0,0) -- (6,6);
\draw[line width=1pt] (0,6) -- (6,0);
\fill (0,0) circle (4pt);
\fill (6,0) circle (4pt);
\fill (0,6) circle (4pt);
\fill (3,3) circle (4pt);
\fill (6,6) circle (4pt);

\fill[blue] (5,5) circle (8pt);
\fill[blue] (1,5) circle (8pt);
\fill[blue] (1,1) circle (8pt);
\fill[blue] (5,1) circle (8pt);
\node[white,scale=0.7] at (5,5) {\bf 2};
\node[white,scale=0.7] at (1,5) {\bf 7};
\node[white,scale=0.7] at (1,1) {\bf 2};
\node[white,scale=0.7] at (5,1) {\bf 7};

\node at (3,1) {\small $(8,\!16)$};
\node at (1.25,3) {\small $(10,\!14)$};
\node at (3,5) {\small $(3,\!7)$};
\node at (4.75,3) {\small $(1,\!9)$};
\node[scale=0.8] at (-0.4,-0.4) {0};
\node[scale=0.8] at (-0.4,3) {1};
\node[scale=0.8] at (3,-0.45) {1};
\node[scale=0.8] at (-0.4,6) {2};
\node[scale=0.8] at (6,-0.45) {2};
\node at (-0.5,4.5) {$q_2$};
\node at (4.5,-0.5) {$q_1$};
\end{tikzpicture}
\vspace*{1mm}
\caption{The left panel shows $-\partial V$ and the right panel shows $\partial U$. Vertices in one graph correspond to regions in the dual graph and their locations produce the labels of the dual regions. In both graphs, the difference in labels between two adjacent regions is the weight (the white number in the blue disk) times the normal to the regions' common edge. Edges in one graph correspond to perpendicular edges in the dual graph, and their weights equal the dual edge's length.}\label{fig:example}
\end{center}
\vspace*{-6mm}
\end{figure}
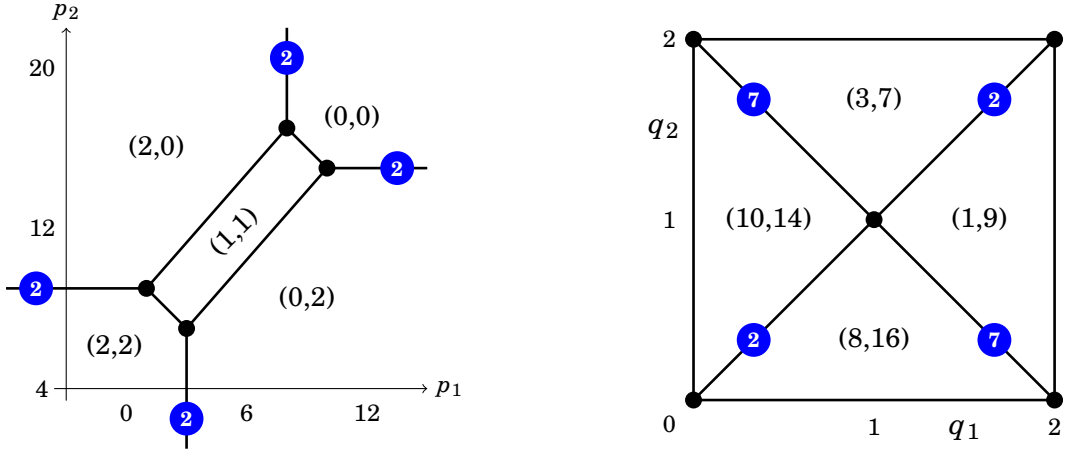

We begin with an example to illustrate the equivalence between \citeauthor{BaldwinKlemperer2019}'s (\citeyear{BaldwinKlemperer2019}) price and quantity complexes and the subdifferential mappings that yield demand and inverse demand (we prove this equivalence more generally in Appendix \ref{App:B}). Consider the set of bundles shown in Figure \ref{fig:discrete} for which a consumer with quasi-linear preferences has the \mbox{indicated} values. The indirect utility is
\begin{equation}\label{discreteIndUtil}
V(\mathbf{p})\,=\,\max(0, 16-2p_1, 24-p_1-p_2, 28-2p_2, 34-2p_1-2p_2)
\end{equation}
and the demand function follows from Roy's identity $\mathbf{Q}(\mathbf{p})=-\partial V(\mathbf{p})$. Since indirect utility is piecewise linear, its subdifferential is piecewise constant. The left panel of Figure~\ref{fig:example}, which is coined the \q{price complex} by \cite{BaldwinKlemperer2019}, shows the price regions where the demand correspondence is single-valued. These regions are separated by a \q{tropical graph,} which contains prices at which demand is multi-valued.\footnote{At these prices, demand equals $-\partial V(p)\cap\field{Z}_{\geq0}^K$ to reflect indivisibility.} \cite{BaldwinKlemperer2019} refer to this tropical graph as the locus of indifference prices (LIP).

The Fenchel dual $U(\mathbf{q})=\min_\mathbf{p}\,V(\mathbf{p})+\langle \mathbf{p}|\mathbf{q}\rangle$ is the lowest concave function everywhere above the discrete valuations, i.e. $U(\mathbf{q})\geq U_\mathbf{q}$ for all $\mathbf{q}\in Q$. It is given by
\begin{equation}\label{discreteUtil}
  U(\mathbf{q})\,=\,\min(14+q_1+9q_2, 14+3q_1+7q_2, 8q_1+16q_2, 10q_1+14q_2)
\end{equation}
for $\mathbf{q} = (q_1,q_2)\in[0,2]^2$.
The projection of the resulting price function $\mathbf{p}(\mathbf{q})=\partial U(\mathbf{q})$ yields the \q{demand complex} shown in the right panel of Figure \ref{fig:example}. As explained in the figure's caption, the graphs in the left and right panels are dual, which is a consequence of the fact that $-\partial V(\mathbf{p})$ and $\partial U(\mathbf{q})$ are inverses in the sense of multi-value mappings, i.e. $\mathbf{q}\in-\partial V(\mathbf{p})$ iff $\mathbf{p}\in \partial U(\mathbf{q})$. Since the graphs contain the same information the optimal demand can be deduced from both.

The price complex in the left panel of Figure \ref{fig:example} coincides with the description of \cite{BaldwinKlemperer2019} but the demand complex in the right panel adds detail to their abstract version. \citeauthor{BaldwinKlemperer2019} (\citeyear[p. 881]{BaldwinKlemperer2019}) write that the price complex \q{shows the actual prices at which bundles are demanded, whereas a demand complex shows only collections of bundles among which the agent is indifferent for some prices,} and that there does not \q{seem to be any simple check of which polyhedral complexes in quantity space correspond to any valuation.}

However, the inverse nature of the price and demand complexes implies they contain the same information. Hence, the demand complex \textit{does} indicate the prices at which a consumer is indifferent, e.g. among $\{\mathpzc{q}_1,\mathpzc{q}_2,\mathpzc{q}_3\}$ when $\mathbf{p}=(8,16)$, between $\{\mathpzc{q}_1,\mathpzc{q}_3\}$ when $\mathbf{p}=(8+\alpha,16-\alpha)$ for $0\leq\alpha\leq 2$, and between $\{\mathpzc{q}_1,\mathpzc{q}_4\}$ when $\mathbf{p}=(10+\alpha,14)$ for $\alpha\geq 0$. Since indifferences define the boundaries of the regions where demand is unique, the actual prices at which bundles are demanded can be inferred from the demand complex.
Second, \citeauthor{BaldwinKlemperer2019}'s (\citeyear[Th. 2.14]{BaldwinKlemperer2019}) criterion whether a price complex stems from utility maximization is that its facet subcomplex satisfies \citeauthor{Mikhalkin2004}'s (\citeyear{Mikhalkin2004}) balancing condition. The same is true for the demand complex. In this case balancedness requires that the vectors normal to the edges in the demand complex sum to zero when weighted by their labels. Going counterclockwise around the $(1,1)$ vertex in the right panel of Figure \ref{fig:example} we have
\begin{displaymath}
  2\,\Bigl(\begin{array}{c}\!\!-1\!\!\! \\ \!\!1\!\!\!\end{array}\Bigr)+7\,\Bigl(\begin{array}{c} \!\!-1\!\!\! \\ \!\!-1\!\!\!\end{array}\Bigr)+2\,\Bigl(\begin{array}{c} \!\!1\!\!\! \\ \!\!-1\!\!\!\end{array}\Bigr)+7\,\Bigl(\begin{array}{c} \!\!1\!\!\! \\ \!\!1\!\!\!\end{array}\Bigr)\,=\,\boldsymbol{0}
\end{displaymath}
The demand complex \textit{is} balanced because $\mathbf{p}(\mathbf{q})$ is conservative, i.e. $\oint_\Gamma \mathbf{p}(\mathbf{q})\cdot d\mathbf{q}=0$ along any closed path $\Gamma$. To summarize, indivisible goods pose no problem for solving the consumer's problem, either in price via Roy's lemma or in quantity space by equating marginal utilities to prices. The reason we can use either space is that the price and demand complexes contain the same information.

Appendix \ref{App:B} extends the results of this section to general price and demand complexes. The price and demand complexes in Figure \ref{fig:example} are examples of normally-labeled polyhedral complexes, see Definitions \ref{polycomp} and \ref{def:labeledpolycomp}, that are equivalent to the vertical projections of the subdifferential of a convex and concave potential respectively (see the Equivalence Theorem \ref{MfromPi}). By ``integrating up'' the price and demand complexes we provide explicit expressions for the polyhedral potentials that generate them (see the Polyhedral Potential Theorem \ref{reversePoly}).\footnote{In the tropical geometry literature, the fact that there exists a tropical polynomial that generates the tropical curve when it is balanced is known as the ``Structure Theorem.'' Our results elucidate that the tropical polynomial is simply a polyhedral potential and that the tropical curve corresponds to points of non-differentiability of the potential.} The potentials are Fenchel duals with subdifferentials that are inverse to each other, which explains the geometric duality between the price and demand complexes noted by \citeauthor{BaldwinKlemperer2019} (see the Geometric Duality Proposition \ref{prop:P-complex1}).

\section{Walrasian Equilibrium}
\label{sec:app}

We consider settings with quasi-linear preferences and indivisible goods, as commonly assumed in the market-design literature. Prominent examples include telecoms bidding for spectrum licenses, e.g. \cite{milgrom2004}, or fisheries trading catch licenses, e.g. \cite{BFG2019}. Before turning to indivisible goods, we first relate Walrasian equilibrium existence in economies with quasi-linear preferences to Fenchel's Duality Theorem, see \citeauthor{Rockafellar1970} (\citeyear[Ch. 31]{Rockafellar1970}). Fenchel's theorem shows that the minimum of a convex potential is equal to the maximum of its concave dual.

\subsection{Equilibrium Existence with Quasi-Linear Preferences}

Consider an exchange economy with consumers $\mathcal{N} = \{1, \ldots, N\}$ and goods $\mathcal{L} = \{1,\ldots, L\}$. The set of feasible allocations is
\begin{displaymath}
F(\omega)\,=\,\bigl\{\,\mathbf{q} \in \field{R}^{NL}_{\geq 0} : \sum_{i\in \mathcal{N}}q_{il} \leq \omega_l\,\,\,\, \forall \,\, l\in \mathcal{L}\,\bigr\}
\end{displaymath}
Let $u_i(\mathbf{q}_i,t_i)=u_i(\mathbf{q}_i)-t_i$ where $u_i(\mathbf{q}_i)$ is consumer $i$'s value for $\mathbf{q}_i$ and $t_i=\langle \mathbf{p}|\mathbf{q}_i-\omega_i\rangle$ is consumer $i$'s transfer. With quasi-linear preferences the indirect utility is $v(\mathbf{p},\omega_i)=\overline{v}_i(\mathbf{p})+\langle \mathbf{p}|\omega_i\rangle$ where $\overline{v}_i(\mathbf{p})$ is the Fenchel dual of $u_i(\mathbf{q}_i)$.
\begin{definition}
A Walrasian equilibrium is a price-allocation pair $(\mathbf{p},\mathbf{q})$ with $\mathbf{p}\in\field{R}_+$ and $\mathbf{q}=\sum_{i}\mathbf{q}_i\in F(\omega)$ such that $v_i(\mathbf{p},\omega_i) = u_i(\mathbf{q}_i)- \langle \mathbf{p}|\mathbf{q}_i-\omega_i\rangle$ for $i\in\mathcal{N}$.
\end{definition}
In words, each consumer $i\in\mathcal{N}$ gets the bundle $\mathbf{q}_i$ that maximizes utility at prices $\mathbf{p}$ and the aggregate demand $\mathbf{q}=\sum_{i}\mathbf{q}_i$ is feasible.
Let $V(\mathbf{p},\omega)=\sum_{i\in\mathcal{N}}v_i(\mathbf{p},\omega_i)$ and $U(\mathbf{q})=\sum_{i\in\mathcal{N}}u_i(\mathbf{q}_i,t_i)=\sum_{i\in\mathcal{N}}u_i(\mathbf{q}_i)$ as the transfers are a wash.
\begin{theorem}[Duality Test]\label{th:dual}
With quasi-linear preferences, $(\mathbf{p},\mathbf{q})$ is a Walrasian equilibrium if and only if $\mathbf{p}\in\field{R}^L_+$ minimizes $V(\mathbf{p},\omega)$ and $\mathbf{q}\in F(\omega)$ maximizes $U(\mathbf{q})$ with
\begin{equation}\label{FenchelTheorem}
  \min_{\mathbf{p}\,\in\,\field{R}^L_+}\,V(\mathbf{p},\omega)\,=\,\max_{\mathbf{q}\,\in\,F(\omega)}\,U(\mathbf{q})
\end{equation}
If the aggregate utility is concave, existence of Walrasian equilibrium follows from Fenchel's Duality Theorem.
\end{theorem}
An easy corollary of Theorem \ref{th:dual} is that with divisible goods, Walrasian equilibrium exists if $u_i(\mathbf{q})$ is concave for $i\in\mathcal{N}$. However, this is not the case for indivisible goods.

\subsection{Indivisible Goods}

The main difficulty with indivisible goods is not the non-differentiability of utility functions but the fact that the aggregate utility function is not necessarily concave even if individual utility functions are. As a result, Walrasian equilibrium may cease to exist.
%Theorem \ref{th:dual} shows that finding the prices that minimize aggregate indirect utility can be done independently from finding the welfare-maximizing allocation. Equilibrium exists if and only if both optimization problems produce the same value.
To illustrate, consider two consumers and two goods, $A$ and $B$, which are both owned by consumer 1. Consumers' values are given in Table \ref{tab:vals}. Notice that the bundle is worth more than the sum of the individual items for consumer 2 (complements) but it is worth less than the sum of the individual items for consumer 1 (substitutes).
The efficient outcome gives both $A$ and $B$ to consumer 2. Can this be sustained by Walrasian prices?

Aggregate indirect utility is given by
\begin{displaymath}
V(p_A,p_B)\,=\,\max(60,30+p_B,50+p_A,p_A+p_B)+\max(0,10-p_A,30-p_B,70-p_A-p_B)
\end{displaymath}
and the unique prices that minimize aggregate indirect utility are $p_A=25$ and $p_B=45$ with $V(25,45)=75$. Aggregate utility is maximized at  $\mathbf{q}_1 = (0,0)$ and $\mathbf{q}_2 = (1,1)$, with $U\big((0,0),(1,1)\big) = 70$.
Hence, the duality test of Theorem \ref{th:dual} is not met and Walrasian equilibrium does not exist.\footnote{When the prices are $p_A=25$ and $p_B=45$, consumers' demands are $\mathbf{q}_1=\{(0,1),(1,0)\}$ and $\mathbf{q}_2=\{(0,0),(1,1)\}$. Their (Minkowski) sum $\{(1,0),(0,1),(2,1),(1,2)\}$ does not include $(1,1)$, i.e. prices are not market clearing. The reason is that at these prices, agent 1 prefers to sell one good rather than both.}

\begin{table}[t]
\begin{center}
\begin{tabular}{c|ccc}
 & $A$ & $B$ & $AB$ \\ \hline
1 & 30 & 50 & 60 \\
2 & 10 & 30 & 70
\end{tabular}
\vspace*{-2mm}
\caption{Consumers' values for (a bundle of) indivisible goods, $A$ and $B$.}\label{tab:vals}
\end{center}
\vspace*{-7mm}
\end{table}

In this example, aggregate utility is not concave. \cite{DKM2001} and \cite{BaldwinKlemperer2019} prove that aggregate utility is concave when a certain unimodularity condition holds for the collection of ``demand types.'' These demand types describe how demand changes in response to price changes and are not tied to specific valuations (like those in Table \ref{tab:vals}). Their result therefore guarantees existence of Walrasian equilibrium for any choice of valuations as long as the unimodularity condition for demand types holds.
If the unimodularity condition is not met, Walrasian equilibrium may still exist for specific valuations. \cite{BaldwinKlemperer2019} provide an argument for how to check for equilibrium existence in this case, which relies on counting the intersections of the tropical curves in the price complex. This is their ``Intersection Count Theorem,'' which underlies the algorithm to check for equilibrium existence in their Appendix A.
The duality test of Theorem \ref{th:dual} provides a simple alternative to check whether Walrasian equilibrium exists.

\section{Conclusions}

We establish the Subgradient Theorem, which is an extension of the well known Gradient Theorem to monotone correspondences. Specifically, we show that any maximal monotone correspondence has a convex or concave potential and, conversely, that the subdifferential of any convex or concave potential defines a maximal conservative monotone correspondence. Further, our Potential Theorem shows how to construct the potential from the correspondence, and our Duality Theorem shows that the inverse of the correspondence is also monotone, maximal and conservative with a potential that is the Fenchel dual of the potential of the original correspondence.

Our results allows us to apply the tools of convex analysis to economies with indivisible goods, generating analogues to classic results including Roy's identity and equating marginal utilities equal to prices. Moreover, it enables a reinterpretation of the important results of \cite{BaldwinKlemperer2019}, couching their duality results and insights within the familiar realm of convex analysis. This allowed us to sharpen their notion of a demand complex to include information on the underlying valuations used to generate demand.

Finally, we introduce a simple test for whether an economy with quasi-linear preferences has a Walrasian equilibrium. If the aggregate utility is concave, equilibrium existence does not require fixed-point arguments but follows from Fenchel's Duality Theorem. With divisible goods, a sufficient condition is that all consumers' utilities are concave. However, when goods are indivisible, aggregate utility is not necessarily concave even if all consumers have concave utility functions. Our duality test in Theorem \ref{th:dual} then provides a simple way to check if Walrasian equilibrium exists.

\newpage

\appendix
\addtolength{\baselineskip}{-1.1mm}

\section{Appendix: Proofs Section 2}
\label{App:A}

We adapt \citeauthor{Aumann1965}'s (\citeyear{Aumann1965}) definition for integrals of multi-valued maps defined on the unit interval. For each $t\in[0,1]$, let $Q(t)$ be a nonempty bounded subset of $\field{R}$. We say that a single-valued function $q:[0,1]\rightarrow\field{R}$ is a \textit{measurable selection} from $Q$ if it is integrable and $m(t)\in Q(t)$ for all $t\in[0,1]$. The Aumann integral of $Q$ is then defined as
\begin{displaymath}
  \int_0^1\,Q(t)dt\,=\,\Bigl\{\int_0^1\,q(t)dt\,:\,q\,\text{is a measurable selection from $Q$}\Bigr\}
\end{displaymath}
Generally, the right side yields a non-empty convex set \citep{Aumann1965}. Our main interest, however, is in \textit{monotone} correspondences, i.e. $(q_2-q_1)(t_2-t_1)\leq 0$ for all $q_1\in Q(t_1)$, $q_2\in Q(t_2)$, $t_1,t_2\in[0,1]$. Throughout we assume that the correspondence is maximal, i.e. its graph is not properly contained in the graph of another monotone correspondence,\footnote{If $M$ is not maximal we replace it with its maximal extension, $\overline{M}$, see footnote \ref{fn:max}. Intuitively, in the univariate case, $\overline{M}$ is such that its graph can be obtained by drawing a curve through the graph of $M$ without lifting one's pen from the paper.} and comment on how results change if it is not. \citeauthor{Kenderov1975} (\citeyear[Th. 2.7]{Kenderov1975}) shows that the set on which a maximal monotone correspondence is multi-valued has measure zero. Different selections from $Q$ are therefore equal almost everywhere and are continuous almost everywhere (as their set of discontinuity points coincides with the set where $Q$ is multi-valued). Since each selection is also bounded, it is Riemann integrable. To summarize, the Aumann integral of a monotone correspondence on $[0,1]$ is unique, i.e. a singleton, and we can use any selection from $Q$ to compute it. Finally, since $Q(t)$ is monotone we have
\begin{displaymath}
  q_0\,\leq\,\int_0^1 Q(t)dt\,\leq\,q_1
\end{displaymath}
for all $q_0\in Q(0)$ and $q_1\in Q(1)$.

Following \cite{Romano1993}, we next extend uniqueness of the Aumann integral to line integrals of maximal monotone correspondences $\mathbf{Q}:C\rightarrow\field{R}^n$ defined on some convex domain $C\subseteq\field{R}^n$.\footnote{If $\mathbf{Q}$ is not maximal then we replace it with its maximal monotone extension $\overline{\mathbf{Q}}$ where, for $\mathbf{p}\in C$, $\overline{\mathbf{Q}}(\mathbf{p})=\{\mathbf{q}\in\field{R}^n:(\mathbf{q}-\mathbf{q}_1)\cdot(\mathbf{p}-\mathbf{p}_1)\geq 0\,\forall\,\mathbf{p}_1\in C,\mathbf{q}_1\in Q(\mathbf{p}_1)\}$. \label{fn:max}} Recall that $\mathbf{Q}$ is \textit{monotone} if and only if
\begin{displaymath}
  (\mathbf{q}_2-\mathbf{q}_1)\cdot(\mathbf{p}_2-\mathbf{p}_1)\,\leq\,0
\end{displaymath}
for all $\mathbf{p}_1,\mathbf{p}_2\in C$, $\mathbf{q}_1\in\mathbf{Q}(\mathbf{p}_1)$, $\mathbf{q}_2\in\mathbf{Q}(\mathbf{q}_2)$. For $\mathbf{q}_1,\mathbf{q}_2\in C$, the projection of $\mathbf{Q}$ along the line segment $\mathbf{p}(t)=(1-t)\mathbf{p}_1+t\mathbf{p}_2$ defines a correspondence on $[0,1]$
\begin{displaymath}
  Q(t)\,=\,\big\{\mathbf{q}(t)\cdot(\mathbf{p}_2-\mathbf{p_1})\,:\,\mathbf{q}(t)\,\in\,\mathbf{Q}(\mathbf{p}(t))\big\}
\end{displaymath}
that is monotone. To see this, note that for all $q_1\in Q(t_1)$, $q_2\in Q(t_2)$ we have
\begin{displaymath}
  (q_2-q_1)(t_2-t_1)=(\mathbf{q}(t_2)-\mathbf{q}(t_1))\cdot(\mathbf{p}_2-\mathbf{p}_1)(t_2-t_1)=(\mathbf{q}(t_2)-\mathbf{q}(t_1))\cdot(\mathbf{p}(t_2)-\mathbf{p}(t_1))\,\leq\,0
\end{displaymath}
where the inequality follows from monotonicity of $\mathbf{Q}$. The line integral of $\mathbf{Q}(\mathbf{p})$ from $\mathbf{p}_1$ to $\mathbf{p}_2$ is thus uniquely defined, i.e. independent on the choice of $\mathbf{q}(t)\in\mathbf{Q}(\mathbf{q}(t))$. Moreover,
\begin{displaymath}
  \mathbf{q}_1\cdot(\mathbf{p}_2-\mathbf{p}_1)\,\geq\,\int_{\mathbf{p}_1}^{\mathbf{p}_2}\mathbf{Q}(\mathbf{p})\cdot d\mathbf{p}\,\geq\,\mathbf{q}_2\cdot(\mathbf{p}_2-\mathbf{p}_1)
\end{displaymath}
for all $\mathbf{q}_1\in\mathbf{Q}(\mathbf{p}_1)$ and $\mathbf{q}_2\in\mathbf{Q}(\mathbf{p}_2)$.

Finally, the integral of $\mathbf{Q}$ along a closed path, $\Gamma$, made out of a finite number of line segments is the sum of the integrals for each of the segments. For $k=1,\ldots,K$ (with $K$ arbitrary), let $\mathbf{p}_k$ denote the start of segment $k$ and let $\mathbf{p}_{k+1}$ denote its end, with $\mathbf{p}_{K+1}\equiv\mathbf{p}_1$ and $\mathbf{q}_{K+1}\equiv\mathbf{q}_1$. We have
\begin{displaymath}
  \sum_{k\,=\,1}^K\mathbf{q}_k\cdot(\mathbf{p}_{k+1}-\mathbf{p}_k)\,\geq\,\oint_{\Gamma}\mathbf{q}(\mathbf{p})\cdot d\mathbf{p}\,\geq\,\sum_{k\,=\,1}^K\mathbf{q}_{k+1}\cdot(\mathbf{p}_{k+1}-\mathbf{p}_k)
\end{displaymath}
If the closed-path integral of $\mathbf{Q}$ along the polyline $\Gamma$ vanishes then the left inequality is the definition of $\mathbf{Q}$ being \textit{cyclically monotone}.\footnote{And the right inequality, which can be rewritten as $\sum_{k=1}^K\mathbf{p}_k\cdot(\mathbf{q}_{k+1}-\mathbf{q}_k)\leq 0$, implies that the inverse of $\mathbf{Q}$ is also cyclically monotone.} \citeauthor{Rockafellar1970} (\citeyear[Th. 24.8, 24.9]{Rockafellar1970}) shows that $\mathbf{Q}$ is cyclically monotone if and only if $\mathbf{Q}(\mathbf{p})\subseteq-\partial V(\mathbf{p})$ for some convex function $V$ with equality when $\mathbf{Q}$ is maximal. \cite{KrishnaMaenner2011} prove that the integral of the subdifferential of a convex function along any closed path\footnote{When we write path we will implicitly assume it is differentiable almost everywhere.} vanishes. Combining these results allows us to define conservative monotone correspondences and state the Subgradient Theorem.\footnote{Note that if the integral of $\mathbf{Q}$ along any closed polyline vanishes then its integral along any closed path vanishes. This might also be shown by approximating an arbitrary closed path by closed polylines with increasingly many segments.}

\medskip

\noindent \textbf{Proof of Theorem \ref{th:subgrad}.} \citeauthor{Rockafellar1970} (\citeyear[Th. 24.9 and Cor. 31.5.2]{Rockafellar1970}) shows that $\partial V$ is maximal cyclically monotone and maximal monotone. \cite{KrishnaMaenner2011} provide an elegant proof that $\partial V$ is conservative. Here we provide a slightly different proof based on the intuition that for any $\mathbf{q}_1,\mathbf{q}_2\in-\partial V(\mathbf{p})$ their difference is normal to the curve that passes through $\mathbf{p}$ almost everywhere. Recall that the directional derivative of $V$ is defined as
\begin{displaymath}
  V'(\mathbf{p};\mathbf{y})\,=\,\lim_{\varepsilon\,\downarrow\,0} \frac{V(\mathbf{p}+\varepsilon\mathbf{y})-V(\mathbf{p})}{\varepsilon}
\end{displaymath}
see \citeauthor{Rockafellar1970} (\citeyear[Sec. 23]{Rockafellar1970}) who shows that $V'(\mathbf{p};\mathbf{y})$ exists and $-V'(\mathbf{p};-\mathbf{y})\leq V'(\mathbf{p};\mathbf{y})$ for all $\mathbf{y}$, see
\citeauthor{Rockafellar1970} (\citeyear[Th. 23.1]{Rockafellar1970}). Moreover, \citeauthor{Rockafellar1970} (\citeyear[Th. 23.2]{Rockafellar1970}) implies
\begin{displaymath}
  -V'(\mathbf{p};-\mathbf{y})\,\leq\,\mathbf{q}\cdot\mathbf{y}\,\leq\,V'(\mathbf{p};-\mathbf{y})
\end{displaymath}
for all $\mathbf{y}$ iff $\mathbf{p}^*$ in $-\partial V(\mathbf{p})$. Hence, for $\mathbf{q}_1,\mathbf{q}_2\in-\partial V(\mathbf{p})$ we have
\begin{displaymath}
  -V'(\mathbf{p};\mathbf{y})-V'(\mathbf{p};-\mathbf{y})\,\leq\,(\mathbf{q}_1-\mathbf{q}_2)\cdot\mathbf{y}\,\leq\,V'(\mathbf{p};\mathbf{y})+V'(\mathbf{p};-\mathbf{y})
\end{displaymath}
Consider a path $\mathbf{p}:[0,1]\rightarrow C$ with $\mathbf{p}(0)=\mathbf{p}(1)$ that is differentiable almost everywhere. Define $\phi(t)=V(\mathbf{p}(t))$, which is regular Lipschitzian, so for almost all $t\in[0,1]$, $\phi(t)$ is differentiable, i.e. $\phi'(t;1)=-\phi'(t;-1)$, or, equivalently, $V'(\mathbf{p}(t);\dot{\mathbf{p}}(t))=-V'(\mathbf{p}(t);-\dot{\mathbf{p}}(t))$. Hence, for almost all $t\in[0,1]$, the above inequality implies $(\mathbf{q}_1-\mathbf{q}_2)\cdot\dot{\mathbf{p}}(t)=0$ for any $q_1,q_2\in-\partial V(\mathbf{p}(t))$, i.e. the difference between two selections from the subdifferential is normal to the path almost everywhere. Hence, $\oint\partial V(\mathbf{p})\cdot d\mathbf{p}=\int_0^1\partial V(\mathbf{p}(t))\cdot\dot{\mathbf{p}}(t)dt$ is independent of the selection from $\partial V(\mathbf{p})$ and since $\partial V(\mathbf{p}(t))\cdot\dot{\mathbf{p}}(t)=V'(\mathbf{p}(t);\dot{\mathbf{p}}(t))=\phi'(t)$ for almost all $t\in[0,1]$, we have $\oint\partial V(\mathbf{p})\cdot d\mathbf{p}=\int_0^1\phi'(t)dt=\phi(1)-\phi(0)=0$, as first shown by \cite{KrishnaMaenner2011}.

For the proof of the converse part, consider the integral of $\mathbf{Q}$ along a closed path, $\Gamma$, made out of a finite number of line segments. For $k=1,\ldots,K$ (with $K$ arbitrary), let $\mathbf{p}_k$ denote the start of segment $k$ and let $\mathbf{p}_{k+1}$ denote its end, with $\mathbf{p}_{K+1}\equiv\mathbf{p}_1$. We have shown in the main text that
\begin{displaymath}
  \sum_{k\,=\,1}^K\mathbf{q}_k\cdot(\mathbf{p}_{k+1}-\mathbf{p}_k)\,\leq\,\oint_{\Gamma}\mathbf{Q}(\mathbf{p})\cdot d\mathbf{p}\,\leq\,\sum_{k\,=\,1}^K\mathbf{q}_{k+1}\cdot(\mathbf{p}_{k+1}-\mathbf{p}_k)
\end{displaymath}
If $\mathbf{Q}$ is conservative, the path integral vanishes and we have
\begin{displaymath}
  \sum_{k\,=\,1}^K\mathbf{q}_k\cdot(\mathbf{p}_{k+1}-\mathbf{p}_k)\,\leq\,0
\end{displaymath}
for all $\mathbf{p}_k\in C$ and $\mathbf{q}_k\in\mathbf{Q}(\mathbf{p}_k)$, which are the inequalities that define cyclical monotonicity of $\mathbf{Q}$, see \citeauthor{Rockafellar1970} (\citeyear[p. 238]{Rockafellar1970}). Indeed, the idea behind choosing a closed polyline is to generate the above inequalities. Hence, $\mathbf{Q}(\mathbf{p})\subseteq-\partial V(\mathbf{p})$ for some convex function $V$, and, if $\mathbf{Q}$ is maximal then $\mathbf{Q}=-\partial V$, see \cite[Th. 24.8 and Th. 24.9]{Rockafellar1970}. (Note that a vanishing closed polyline integral of $M$ also implies
\begin{displaymath}
  0\,\leq\,\sum_{k\,=\,1}^K\mathbf{q}_{k+1}\cdot(\mathbf{p}_{k+1}-\mathbf{p}_k)=-\sum_{k=1}^K\mathbf{p}_k\cdot(\mathbf{q}_{k+1}-\mathbf{q}_k)
\end{displaymath}
which implies that the inverse $\mathbf{P}$ of $\mathbf{Q}$ is also cyclically monotone and, hence, $\mathbf{P}\subseteq\partial U$ for some concave potential $U$ with equality when $\mathbf{P}$ is maximal.) \hfill$\blacksquare$

An easy corollary is that a monotone correspondence is cyclically monotone if and only if it is conservative. (Note that every cyclically monotone correspondence is monotone, e.g. use $K=2$ in the definition of cyclical monotonicity.)

\medskip

\noindent \textbf{Proof of Theorem \ref{th:potential}.} Let $\mathbf{p}_1\in C$ and let $\Gamma(\mathbf{p}_1,\mathbf{p}_2)$ denote the line from $\mathbf{p}_1$ to $\mathbf{p}_2$. From definition of the potential \eqref{potential} and the proof of Theorem \ref{th:subgrad} we have
\begin{displaymath}
  V(\mathbf{p}_2)\,=\,V(\mathbf{p}_1)-\int_{\Gamma(\mathbf{p}_1,\mathbf{p}_2)}\,\mathbf{Q}(\mathbf{p})\cdot d\mathbf{p}\,\geq\,V(\mathbf{p}_1)-\mathbf{q}_1\cdot(\mathbf{p}_2-\mathbf{p}_1)
\end{displaymath}
for all $\mathbf{p}_2\in C$, $\mathbf{q}_1\in\mathbf{Q}(\mathbf{p}_1)$. Hence, $\mathbf{q_1}\in-\partial V(\mathbf{p}_1)$ for all $\mathbf{p}_1\in C$ and $\mathbf{q}_1\in\mathbf{Q}(\mathbf{p}_1)$, i.e. $\mathbf{Q}\subseteq-\partial V$ (with equality when $\mathbf{Q}$ is maximal, see above).\hfill$\blacksquare$

\section{Appendix: Proofs Section 3}
\label{App:B}

A polyhedral convex function $V:\field{R}^n\rightarrow\field{R}^n$ is a piecewise linear function of the type $V(\mathbf{p})=\max_{k\,\in\,K}(-\boldsymbol{\ell}_k\cdot\mathbf{p}+c_k)$ for some $\boldsymbol{\ell}_k\in\field{R}^n$, $c_k\in\field{R}$, and some finite set $K$, i.e. $V$ is finitely generated. The vertical projection of the graph of $V$ yields a polyhedral subdivision of $\field{R}^n$, i.e. a collection of polyhedra glued together along their faces that covers $\field{R}^n$.
\begin{definition}\label{polycomp} A polyhedral complex, $\Pi\subseteq\field{R}^n$, is a finite collection of {\em cells} $\mathscr{C}\subseteq\field{R}^n$ such that:
\vspace*{-3mm}
\begin{itemize}[noitemsep]
\item[(i)] if $\mathscr{C}\in\Pi$ then $\mathscr{C}$ is a polyhedron\footnote{Recall that a polyhedron is the intersection of finitely many half-spaces $\{\mathbf{p}\in\field{R}^n\,:\,\mathbf{p}\cdot\mathbf{q}\leq c\}$ for some $\mathbf{q}\in\field{R}^n$ and $c\in\field{R}$.} and any face of $\mathscr{C}$ is also in $\Pi$;
\item[(ii)] if $\mathscr{C},\mathscr{C}'\in\Pi$ and $\mathscr{C}\cap\mathscr{C}'\neq\emptyset$ then $\mathscr{C}\cap\mathscr{C}'$ is a face of both $\mathscr{C}$ and $\mathscr{C}'$.
\end{itemize}
\vspace*{-3mm}
\noindent A $k$-cell is a cell of dimension $k$: $(n-1)$-cells, $1$-cells, and 0-cells are also referred to as {\em facets}, {\em edges}, and {\em vertices} respectively. The boundary of a $k$-cell is a union of $(k-1)$-cells. If the union of $n$-cells of $\Pi$ cover $\field{R}^n$ then $\Pi$ is called a polyhedral subdivision of $\field{R}^n$.
The dimension of $\Pi$ is the maximal dimension of its cells.
\end{definition}
\begin{definition}\label{def:labeledpolycomp} The pair $\{\Pi,\boldsymbol{\ell}\}$ defines a {\em normally labeled} polyhedral complex if $\Pi\subseteq\field{R}^n$ is a polyhedral complex and
$\boldsymbol{\ell}(\mathscr{C}_n)\in\field{R}^n$ a set of labels, one for each $n$-cell $\mathscr{C}_n$, such that for all adjacent $n$-cells $\mathscr{C}_n\neq\mathscr{C}'_n$, $\boldsymbol{\ell}(\mathscr{C}'_n)-\boldsymbol{\ell}(\mathscr{C}_n)$ equals a weight, $w(\mathscr{C}_{n-1})\in\field{R}_{>0}$, times a unit vector that points from $\mathscr{C}_n$ to $\mathscr{C}'_n$ and is normal to the facet $\mathscr{C}_{n-1}=\mathscr{C}_n\cap\mathscr{C}'_n$.
\end{definition}
A full-dimensional cell in the graph of a polyhedral convex function $V$ corresponds to a region where $V$ is linear, i.e. given by $-\boldsymbol{\ell}_k\cdot\mathbf{p}+c_k$ for some $k\in K$. If we label the (vertical) projection of this cell with the relevant $\boldsymbol{\ell}_k$ then this results in a normally labeled subdivision. To see this, suppose $\mathbf{x}$ and $\mathbf{y}$ are two different elements of a facet that separates two adjacent full-dimensional regions labeled $\boldsymbol{\ell}_k$ and $\boldsymbol{\ell}_{k'}$. Then $-\boldsymbol{\ell}_k\cdot\mathbf{x}+c_k=-\boldsymbol{\ell}_{k'}\cdot\mathbf{x}+c_{k'}$ and
$-\boldsymbol{\ell}_k\cdot\mathbf{y}+c_k=-\boldsymbol{\ell}_{k'}\cdot\mathbf{y}+c_{k'}$. Taking differences yields $-(\boldsymbol{\ell}_{k}-\boldsymbol{\ell}_{k'})\cdot(\mathbf{x}-\mathbf{y})=0$, i.e. the difference in labels is perpendicular to a vector parallel to the facet. Conversely, based on the Potential Theorem \ref{th:potential}, we show in Proposition \ref{reversePoly} below that the normally labeled complex can be \q{integrated up} to a polyhedral convex function. This requires conservativeness, which we show by establishing equivalence between normally labeled subdivisions and subdifferential mappings.

Recall that $\mathbf{n}\in\partial V(\mathbf{p})$ iff $(\mathbf{n},-1)$ defines a supporting hyperplane to the graph of $V$ at $(\mathbf{p},V(\mathbf{p}))$. Hence, for any $k$-dimensional cell in the graph, $\mathbf{Q}(\mathbf{p})=-\partial V(\mathbf{p})$ is constant for $\mathbf{p}$ in the relative interior of the cell and of dimension $n-k$, i.e. the number of independent normals to the affine span of the $k$-dimensional cell. In particular, for $\mathbf{p}$ in the relative interior of a full-dimensional cell, $\mathbf{Q}(\mathbf{p})$ is a singleton, i.e. $\mathbf{Q}(\mathbf{p})=\boldsymbol{\ell}_k$ for some $k\in K$. Hence, to the subdifferential mapping $\mathbf{Q}$ we can associate a normally labeled subdivision, $\{\Pi_\mathbf{Q},\boldsymbol{\ell}_\mathbf{Q}\}$, of $\field{R}^n$ as follows: let $\Pi_\mathbf{Q}$ to be the collection of cells on which $\mathbf{Q}$ and, hence, its dimension are constant, i.e. $\mathscr{C}_k=\text{cl}\{\mathbf{p}\in\field{R}^n:\dim(\mathbf{Q}(\mathbf{p}))=n-k\}$ for $k=0,\ldots,n$, and let $\boldsymbol{\ell}_\mathbf{Q}(\mathscr{C}_n)=\mathbf{Q}(\text{int}(\mathscr{C}_n))$.\footnote{Here \q{$\text{cl}$} denotes topological closure and \q{$\text{int}$} relative interior.}

\begin{figure}[t]
\begin{center}
\begin{tikzpicture}[scale=0.7]
\draw[line width=1pt] (-2,-3) -- (0,0) -- (3,-1);
\draw[line width=1pt] (2,4) -- (0,0) -- (-4,2);
\draw[loosely dotted,very thick] (-3,1.8) arc (140:72:4cm);
\fill[black] (0,0) circle (8pt);
\node[below,scale=0.8pt] at (0.3,-0.25) {\small $\mathscr{C}_{n-2}$};

\node[below,scale=0.8] at (-2,-3) {\small $\mathscr{C}_{n-1}^1$};
\node[right,scale=0.8]  at (3,-1) {\small $\mathscr{C}_{n-1}^2$};
\node[above,scale=0.8] at (2,4) {\small $\mathscr{C}_{n-1}^3$};
\node[left,scale=0.8] at (-4,2) {\small $\mathscr{C}_{n-1}^K$};
\draw[->] (-1,-1.5) -- (0,-1.5-2/3) node [right] {$\mathbf{n}_1$};
\fill[blue] (-1,-1.5) circle (8pt);
\node[white,scale=0.65] at (-1,-1.5) {$\mathbf{w}_1$};
\draw[->] (1.5,-0.5) -- (1.5+1/3,0.5) node [right] {$\mathbf{n}_2$};
\fill[blue] (1.5,-0.5) circle (8pt);
\node[white,scale=0.65] at (1.5,-0.5) {$\mathbf{w}_2$};
\draw[->] (0.75,1.5) -- (-0.25,2) node [left] {$\mathbf{n}_3$};
\fill[blue] (0.75,1.5) circle (8pt);
\node[white,scale=0.65] at (0.75,1.5) {$\mathbf{w}_3$};
\draw[->] (-1.5,0.75) -- (-2,-0.25) node [below] {$\mathbf{n}_{\!K}$};
\fill[blue] (-1.5,0.75) circle (8pt);
\node[white,scale=0.65] at (-1.5,0.75) {$\mathbf{w}_{\!K}$};
\node[scale=1.2] at (2,-2.5) {$\boldsymbol{\ell}_1$};
\node[scale=1.2] at (3,2) {$\boldsymbol{\ell}_2$};
\node[scale=1.2] at (-3.5,-1) {$\boldsymbol{\ell}_K$};
\end{tikzpicture}
\caption{Equivalence between balancedness and normal labeling.}\label{fig:balanced}
\end{center}
\vspace*{-5mm}
\end{figure}
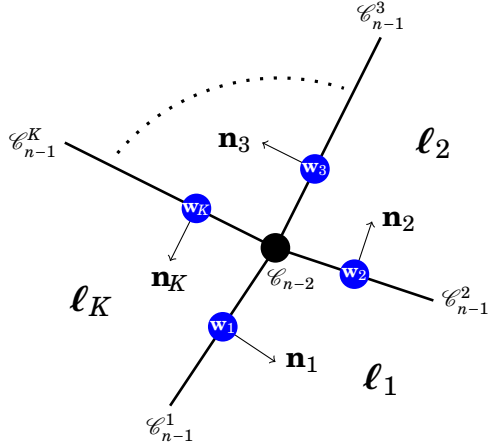

Conversely, we can \q{lift} any normally labeled polyhedral subdivision $\{\Pi,\boldsymbol{\ell}\}$ of $\field{R}^n$ to define the correspondence $\mathbf{Q}_{\Pi,\ell}(\mathbf{p})=\text{co}\{\boldsymbol{\ell}(\mathscr{C}_n):\mathbf{p}\in\mathscr{C}_n\subseteq\Pi\}$, where \q{co} denotes the convex hull. This correspondence is maximal. We next show it is also monotone and conservative, i.e. it is a subdifferential mapping by Theorem \ref{th:subgrad}. Definition \ref{def:labeledpolycomp} implies the $i$-th label is non-decreasing in the $i$-th coordinate: $\ell_{2,i}\leq\ell_{1,i}$ if and only if $p_{2,i}\geq p_{1,i}$, or $(\mathbf{p}_2-\mathbf{p}_1)\cdot(\boldsymbol{\ell}_2-\boldsymbol{\ell}_1)\leq 0$. Hence $\mathbf{Q}_{\Pi,\ell}$ is monotone. To show conservativeness, we first note that the facet subcomplex of a normally labeled complex is  \textit{balanced} in the sense of \cite{Mikhalkin2004}. Consider the configuration in Figure \ref{fig:balanced} where $K$ weighted facets with unit normals $\mathbf{n}_k$ intersect at an $(n-2)$-dimensional cell, $\mathscr{C}_{n-2}$, and the $K$ adjacent regions are labeled $\boldsymbol{\ell}_k$ for $k=1,\ldots,K$. Normal labeling requires e.g. $\boldsymbol{\ell}_2=\boldsymbol{\ell}_1+w_2\mathbf{n}_2$ and $\boldsymbol{\ell}_3=\boldsymbol{\ell}_2+w_3\mathbf{n}_3$ etc. If we make a full (counterclockwise) circle we get
\begin{displaymath}
  \boldsymbol{\ell}_1\,=\,\boldsymbol{\ell}_K+w_1\mathbf{n}_1\,=\,\boldsymbol{\ell}_1+\sum_{k\,=\,1}^Kw_k\mathbf{n}_k
\end{displaymath}
i.e. normal labeling is equivalent to \citeauthor{Mikhalkin2004}'s (\citeyear{Mikhalkin2004}) balancing condition
\begin{displaymath}
  \sum_{k\,=\,1}^K\,w_k\mathbf{n}_k\,=\,\boldsymbol{0}
\end{displaymath}
Since $\mathbf{Q}_{\Pi,\ell}$ is monotone, its integral along a closed path $\Gamma$ is independent of the selection from $\mathbf{Q}_{\Pi,\ell}$. Let $\Gamma$ denote any closed path around $\mathscr{C}_{n-2}$ in Figure \ref{fig:balanced} above. Let $\mathbf{p}^i_{k}$ and $\mathbf{p}^o_{k}$ denote the entry and exit points of $\Gamma$ for region $k$, and let $\mathbf{p}_0$ denote any vector in $\mathscr{C}_{n-2}$. We do not require $\mathbf{p}^i_{k+1}=\mathbf{p}^o_{k}$, i.e. the path may travel through the facet $\mathscr{C}_{n-1}^{k+1}$ where $\mathbf{Q}_{\Pi,\ell}$ is multi-valued. Using the convention $K+1\equiv 1$, we have
\begin{eqnarray*}
  \oint_{\Gamma}\,\mathbf{Q}_{\Pi,\ell}(\mathbf{p})\cdot d\mathbf{p} &=& \sum_{k\,=\,1}^K\boldsymbol{\ell}_{k}\cdot(\mathbf{p}^o_{k}-\mathbf{p}^i_{k})+\sum_{k\,=\,1}^K\boldsymbol{\ell}_{k}\cdot(\mathbf{p}^i_{k+1}-\mathbf{p}^o_{k})\\[1mm]
  &=& \sum_{k\,=\,1}^K\mathbf{p}^i_{k+1}\cdot(\boldsymbol{\ell}_{k}-\boldsymbol{\ell}_{k+1}) \\[1mm]
  &=& \mathbf{p}_0\cdot\sum_{k\,=\,1}^K w_k\mathbf{n}_k
\end{eqnarray*}
where in the first line we used that we can take any selection from $\mathbf{Q}_{\Pi,\ell}$ on the facets, and in the third line we used $\boldsymbol{\ell}_{k}-\boldsymbol{\ell}_{k+1}=w_{k+1}\mathbf{n}_{k+1}$ and $(\mathbf{p}^i_{k+1}-\mathbf{p}_0)\cdot\mathbf{n}_{k+1}=0$ for $k=1,\ldots,K$. Hence, conservativeness of $\mathbf{Q}_{\Pi,\ell}$ follows from normal labeling of $\Pi$.\footnote{Since we can perturb the polyhedral complex, including $\mathbf{p}_0$, without affecting its topological structure, including its labels, conservativeness of the correspondence $\mathbf{Q}$ is both necessary and sufficient for the complex to be normally labeled (or, equivalently, for its facet complex to be balanced).} To summarize:
\begin{theorem}[Equivalence]\label{MfromPi}
There exists a bijection between normally labeled polyhedral subdivisions of $\field{R}^n$ and subdifferential mappings of polyhedral convex functions.
\end{theorem}

\noindent \textbf{Proof.} %\textbf{Proof of Theorem \ref{MfromPi}.}
An expanded version of the statement of Theorem \ref{MfromPi} is as follows:
\begin{itemize}
\item[(1)] If $\{\Pi,\boldsymbol{\ell}\}$ is a normally labeled polyhedral subdivision of $\field{R}^n$ then $\mathbf{Q}(\mathbf{p})=\text{\em co}(\boldsymbol{\ell}(\mathscr{C}_n):\mathbf{p}\in\mathscr{C}_n)$ is a maximal, monotone decreasing, and conservative correspondence, i.e. $\mathbf{Q}=-\partial V$ for some polyhedral convex function $V$.
\item[(2)] Conversely, suppose $\mathbf{Q}=-\partial V$ for some polyhedral convex function $V$ with domain $\field{R}^n$. Let $\Pi$ denote the collection of cells $\mathscr{C}_k=\text{\em cl}(\mathbf{x}\in\field{R}^n:\dim(\mathbf{Q}(\mathbf{p}))=n-k)$ for $k=0,\ldots,n$, and let $\mathbf{Q}(\mathscr{C}_n)$ be the label for $\mathscr{C}_n$. Then $\{\Pi,\boldsymbol{\ell}\}$ defines a normally labeled polyhedral subdivision of $\field{R}^n$.
\end{itemize}
Maximality of $\mathbf{Q}$ is obvious and we already showed $\mathbf{Q}$ is monotone decreasing and conservative, which proves (1). To prove the converse part (2),
recall that the (projected) graph of a polyhedral convex function $V$ forms a subdivision of its domain, see e.g. \citeauthor{ContMath2013}(\citeyear[p. 185]{ContMath2013}). Moreover, $\mathbf{n}\in\partial V(\mathbf{p})$ iff $(\mathbf{n},-1)$ defines a supporting hyperplane to the graph of $V$ at $(\mathbf{p},V(\mathbf{p}))$. Hence, for any $k$-dimensional cell in the graph, $\mathbf{Q}(\mathbf{p}) =-\partial V(\mathbf{p})$ is constant on its interior and of dimension $n-k$, i.e. the number of independent normals to the affine span of the $k$-dimensional cell. Finally, since $\mathbf{Q}=-\partial V$ is conservative, balancedness of the facet complex of $\Pi$ holds, or, equivalently, $\{\Pi,\boldsymbol{\ell}\}$ is normally labeled.\hfill$\blacksquare$\\[0.5mm]

\noindent If $V$ is a polyhedral function then so is its Fenchel dual $U$. Recall that $\mathbf{P}=\partial U$ is the inverse of $\mathbf{Q}=-\partial V$.  The range of $\mathbf{Q}$ is the finite polytope $\text{co}(\boldsymbol{\ell})$, which is the domain of $\mathbf{P}$.\footnote{Since the graphs of $\mathbf{Q}=-\partial V$ and $\mathbf{P}=\partial U$ are homeomorphic to $\field{R}^n$, see \citeauthor{Rockafellar1970} (\citeyear[Cor. 31.5.1]{Rockafellar1970}), either the domain of $V$ or the domain of $U$ (or both) will be $\field{R}^n$. Here we choose, without loss of generality, $\text{dom}(V)=\field{R}^n$.} On this domain, $\mathbf{P}$ defines a normally labeled subdivision $\{\Pi^*,\boldsymbol{\ell}^*\}$ that is dual to the normally labeled polyhedral subdivision $\{\Pi,\boldsymbol{\ell}\}$ of $\field{R}^n$ defined by $\mathbf{Q}$.

\begin{proposition}[Geometric Duality]\label{prop:P-complex1} The normally labeled subdivisions $\{\Pi,\boldsymbol{\ell}\}$ and $\{\Pi^*,\boldsymbol{\ell}^*\}$ are dual, i.e. there exists a bijection between cells $\mathscr{C}\subseteq\Pi$ and $\mathscr{C}^*\subseteq\Pi^*$ given by $\mathscr{C}^*_{n-k}=\mathbf{Q}(\text{\em int}(\mathscr{C}_{k}))$ and $\mathscr{C}_k=\mathbf{P}(\text{\em int}(\mathscr{C}^*_{n-k}))$ for $k=0,\ldots,n$, and $\boldsymbol{\ell}$ and $\boldsymbol{\ell}^*$ are given by the 0-cells of $\Pi^*$ and $\Pi$ respectively. The cells $\mathscr{C}_{k}$ and $\mathscr{C}^*_{n-k}$ are orthogonal for $k=0,\ldots,n$.
\end{proposition}

\noindent \textbf{Proof.} %\textbf{Proof of Proposition \ref{prop:P-complex1}.}
The polyhedral complexes $\Pi$ and $\Pi^*$ correspond to the projected graphs of $V$ and $U$ respectively. The dual nature of $V(\mathbf{p})$ and $U(\mathbf{q})$ implies a bijective correspondence $\mathscr{C}\leftrightarrows\mathscr{C}^*$ between cells in $\Pi$ and $\Pi^*$ given by
\begin{displaymath}
  \mathscr{C}^*\,=\,\big\{\mathbf{q}\in\text{co}(\boldsymbol{\ell})\,:\,\mathbf{p}\cdot\mathbf{q}\,=\,V(\mathbf{p})+U(\mathbf{q})\,\,\forall\,\mathbf{q}\in\mathscr{C}\big\}
\end{displaymath}
and
\begin{displaymath}
  \mathscr{C}\,=\,\big\{\mathbf{p}\in\field{R}^n\,:\,\mathbf{p}\cdot\mathbf{q}\,=\,V(\mathbf{p})+U(\mathbf{q})\,\,\forall\,\mathbf{q}\in\mathscr{C}^*\big\}
\end{displaymath}
where $\dim(\mathscr{C})+\dim(\mathscr{C}^*)=n$, see e.g. \citeauthor{ContMath2013}(\citeyear[p. 185]{ContMath2013}). The equality $\mathbf{p}\cdot\mathbf{q}=V(\mathbf{p})+U(\mathbf{q})$ holds for all $\mathbf{p}\in\mathscr{C}$ iff $\mathbf{q}\in-\partial V(\mathbf{p})=\mathbf{Q}(\mathbf{p})$, see \citeauthor{Rockafellar1970} (1970, Th. 23.5). Since $\mathbf{Q}$ is constant on the interior of $\mathscr{C}$ this implies that $\mathscr{C}^*=\mathbf{Q}(\text{int}(\mathscr{C}))$. The converse reasoning holds as well, i.e. $\mathbf{p}\cdot\mathbf{q}=V(\mathbf{p})+U(\mathbf{q})$ holds for all $\mathbf{q}\in\mathscr{C}^*$ iff $\mathbf{q}\in\mathbf{P}(\mathbf{q})$, from which it follows that $\mathscr{C}=\mathbf{P}(\text{int}(\mathscr{C}^*))$.

The labels of $\Pi^*$ are given by the values of $\mathbf{P}$ on the interior of its full-dimensional sets, which are the images of the 0-cells of $\Pi$ under $\mathbf{Q}$. Let $\mathbf{p}$ belong to some 0-cell of $\Pi$ then the resulting label for $\Pi^*$ is $\mathbf{P}(\text{int}(\mathbf{Q}(\mathbf{p})))=\mathbf{p}$. Similarly for the labels of $\Pi$.

Finally, for any $\mathbf{q}$ that lies in the facet dividing adjacent full-dimensional cells of $\Pi^*$ with labels $\boldsymbol{\ell}^*_1,\boldsymbol{\ell}^*_2$, we have
\begin{displaymath}
  \mathbf{q}\cdot(\boldsymbol{\ell}^*_2-\boldsymbol{\ell}^*_1)\,=\,V(\boldsymbol{\ell}^*_2)-V(\boldsymbol{\ell}^*_1)
\end{displaymath}
Hence, for any $\mathbf{q}_1\neq\mathbf{q}_2$ that lie in this facet we have $(\mathbf{q}_2-\mathbf{q}_1)\cdot(\boldsymbol{\ell}^*_2-\boldsymbol{\ell}^*_1)=0$, i.e. the difference in labels is normal to the facet. An analogous argument applies to $\Pi$.

Here we show orthogonality. For $\mathbf{p}_1,\mathbf{p}_2\in\mathscr{C}_k$ and $\mathbf{q}_1,\mathbf{q}_2\in\mathscr{C}^*_{n-k}$ we have\footnote{Since $\mathbf{q}_i\in-\partial V(\mathbf{p}_i)$ and $\mathbf{p}_i\in\partial U(\mathbf{q}_i)$ for $i=1,2$, see \citeauthor{Rockafellar1970} (\citeyear[Th. 23.5]{Rockafellar1970}).} $\mathbf{p}_1\cdot\mathbf{q}_1=V(\mathbf{p}_1)+U(\mathbf{q}_1)$ and $\mathbf{p}_2\cdot\mathbf{q}_1=V(\mathbf{p}_2)+U(\mathbf{q}_1)$, so taking differences,
$(\mathbf{p}_2-\mathbf{p}_1)\cdot\mathbf{q}_1=V(\mathbf{p}_1)-V(\mathbf{p}_2)$. But also $(\mathbf{p}_2-\mathbf{p}_1)\cdot\mathbf{q}_2=V(\mathbf{p}_1)-V(\mathbf{p}_2)$. Taking differences one more time, $(\mathbf{p}_2-\mathbf{p}_1)\cdot(\mathbf{q}_2-\mathbf{q}_1)=0$, i.e. the cells $\mathscr{C}_k$ and $\mathscr{C}^*_{n-k}$ are orthogonal.\hfill$\blacksquare$\\[0.5mm]

Since duality is a consequence of the inverse nature of $\mathbf{Q}$ and $\mathbf{P}$, another interpretation is in terms of horizontal and vertical projections. The cells on which $\mathbf{Q}$ is constant produces $\Pi$. This is the \q{vertical projection} of $\mathbf{Q}$. And, by Proposition \ref{prop:P-complex1}, the value of $\mathbf{Q}$ on the cells where it is constant produces $\Pi^*$. This is the \q{horizontal projection} of $\Pi$. Conversely, the vertical projection of $\mathbf{P}$ yields $\Pi^*$ and its horizontal projection yields $\Pi$. To summarize, either $\mathbf{Q}$ or $\mathbf{P}$ by itself can be used to define both complexes, which is not surprising since they are inverses so all information contained in $\mathbf{P}$ must already be encoded in $\mathbf{Q}$, and vice versa.\footnote{Similar to the construction of $\mathbf{Q}_{\Pi,\ell}$ from $\{\Pi,\boldsymbol{\ell}\}$ defined above, $\mathbf{P}$ can be obtained from $\{\Pi^*,\boldsymbol{\ell}^*\}$. The only difference is that $\mathbf{P}$ has finite domain $\text{co}(\boldsymbol{\ell})$ outside which its potential $U=\infty$. As a result, $\mathbf{P}(\mathbf{q})$ is equal to an unbounded polyhedral set when $\mathbf{q}$ belongs to the boundary of the domain.}

Finally, using Theorem \ref{th:potential} we can be more specific about the concave potentials $V$ and $U$.
\begin{theorem}[Polyhedral Potentials]\label{reversePoly}
Let $\{\Pi,\boldsymbol{\ell}\}$ be a normally labeled polyhedral subdivision of $\field{R}^n$ with $n$-cells, $\mathscr{C}_n^k$, labeled by $\boldsymbol{\ell}_k\in\field{R}^n$ for $k=1,\ldots,|\mathscr{C}_n|$, and 0-cells, $\mathscr{C}_0^k$, located at $\boldsymbol{\ell}^*_k$ for $k=1,\ldots,|\mathscr{C}_0|$. Define $V:\field{R}^n\rightarrow\field{R}$ as follows
\begin{displaymath}\label{DC}
V(\mathbf{p})\,=\,\max_{k\,=\,1,\ldots,|\mathscr{C}_n|}\,(\boldsymbol{\ell}_k\cdot\mathbf{p}+c_k)
\end{displaymath}
where $c_i-c_j=\boldsymbol{\ell}^*\cdot(\boldsymbol{\ell}_j-\boldsymbol{\ell}_i)$ if $\mathscr{C}_n^i\cap\mathscr{C}_n^j$ is non-empty, and $\boldsymbol{\ell}^*$ is any 0-cell in this intersection. Furthermore, define $U:\text{\em co}(\boldsymbol{\ell})\rightarrow\field{R}$ as follows\footnote{The dual potential $U$ is assumed to be $\infty$ outside of $\text{co}(\boldsymbol{\ell})$.}
\begin{displaymath}\label{DDC}
U(\mathbf{q})\,=\,\max_{k\,=\,1,\ldots,|\mathscr{C}_0|}\,(\boldsymbol{\ell}^*_k\cdot\mathbf{q}+c^*_k)
\end{displaymath}
where $c^*_i-c^*_j=\boldsymbol{\ell}\cdot(\boldsymbol{\ell}^*_j-\boldsymbol{\ell}^*_i)$ if there is an $n$-cell containing both $\boldsymbol{\ell}^*_i$ and $\boldsymbol{\ell}^*_j$, and $\boldsymbol{\ell}$ is the label of any such $n$-cell. Then the vertical projections of $\mathbf{Q}(\mathbf{p})=-\partial V(\mathbf{p})$ and $\mathbf{P}(\mathbf{q})=\partial U(\mathbf{q})$  yield $\{\Pi,\boldsymbol{\ell}\}$ and $\{\Pi^*,\boldsymbol{\ell}^*\}$ respectively.
\end{theorem}

\noindent \textbf{Proof.} %\textbf{Proof of Proposition \ref{reversePoly}.}
The constructed functions are continuous, piecewise linear, and concave, i.e. they are polyhedral concave functions. To show that $-\partial V(\mathbf{p})=\boldsymbol{\ell}_i$ for $\mathbf{p}\in\mathscr{C}_n^i$, we need $\boldsymbol{\ell}_i\cdot\mathbf{p}+c_i\leq\boldsymbol{\ell}_j\cdot\mathbf{p}+c_j$, or, equivalently,
\begin{displaymath}
  (\mathbf{p}-\boldsymbol{\ell}^*)\cdot(\boldsymbol{\ell}_i-\boldsymbol{\ell}_j)\,\leq\,0
\end{displaymath}
for all $j=1,\ldots,|\mathscr{C}_n|$. In light of Definition \ref{def:labeledpolycomp}, normal labeling of $\Pi$ implies that the $i$-th label is non-increasing in the $i$-th coordinate, i.e.
$\ell_{i}\leq\ell'_{i}$ iff $p_{i}\geq p'_{i}$, or $(\mathbf{p}-\mathbf{p}')\cdot(\boldsymbol{\ell}-\boldsymbol{\ell}')\geq 0$ for any $\mathbf{p}\in\mathscr{C}_n$ and $\mathbf{p}\in\mathscr{C}'_n$. The result follows by taking $\mathscr{C}_n=\mathscr{C}_n^i$, $\mathscr{C}'_n=\mathscr{C}_n^j$, and $\mathbf{p}'=\boldsymbol{\ell}^*\in\mathscr{C}_n^j$. A similar argument applies to $\Pi^*$.\hfill$\blacksquare$

\section{Appendix: Proofs Section 4}
\label{App:C}

Let $f(\mathbf{q})$ be a concave function and $g(\mathbf{q})$ be a convex function. Their Fenchel \mbox{duals} $f^*(\mathbf{p})=\max_\mathbf{q}\,f(\mathbf{q})-\langle\mathbf{p}|\mathbf{x}\rangle$ and $g^*(\mathbf{p})=\min_\mathbf{q}\,g(\mathbf{q})+\langle\mathbf{p}|\mathbf{x}\rangle$ are convex and concave functions respectively. Fenchel's Duality Theorem impies that
\begin{equation}\label{appC:FDT}
\max_\mathbf{q}\,f(\mathbf{q})-g(\mathbf{q})\,=\,\min_\mathbf{p}\,f^*(\mathbf{p})-g^*(\mathbf{p})
\end{equation}
when, e.g., the domains of $f$ and $g$ and the domains of $f^*$ and $g^*$ are the same, see \citeauthor{Rockafellar1970} (\citeyear[Ch. 31]{Rockafellar1970}). For our purposes, we assume $f$ and $g$ are defined over some feasible set $F(\omega)\subset\field{R}^L_{\geq 0}$ and that $g=0$ on this domain. The domains of $f^*$ and $g^*$ are $\field{R}^L_+$ and it is readily verified that $g^*=0$. Hence, \eqref{appC:FDT} simplifies to
\begin{equation}\label{appC:FDT2}
\max_{\mathbf{q}\,\in\,F(\omega)}\,U(\mathbf{q})\,=\,\min_{\mathbf{p}\,\in\,\field{R}^L_+}\,V(\mathbf{p})
\end{equation}
where we defined $U(\mathbf{q})=f(\mathbf{q})$ and $V(\mathbf{p})=g(\mathbf{p})$.

\medskip

\noindent \textbf{Proof of Theorem \ref{th:dual}.} Following \cite{Goeree2023}, we define the economy's potential $Y(\mathbf{p},\mathbf{q},\omega)=U(\mathbf{q})-V(\mathbf{p},\omega)$, i.e.
\begin{eqnarray*}
  Y(\mathbf{p},\mathbf{q},\omega) &=& \sum_{i\,\in\,\mathcal{N}}\,u_i(q_i)-v_i(\mathbf{p},\omega_i) \\[1mm]
  &=& \sum_{i\,\in\,\mathcal{N}}\,u_i(q_i)-\langle\mathbf{p}|\mathbf{q}_i-\omega_i\rangle-v_i(\mathbf{p},\omega_i)
\end{eqnarray*}
where the second line follows since the transfers are a wash. Note that each term in the sum is non-positive so $Y:\field{R}^L_+\times F(\omega)\rightarrow\field{R}_{\leq0}$. Hence, if $Y(\mathbf{p},\mathbf{q},\omega)=0$ then each term in the sum is nil and $(\mathbf{p},\mathbf{q})$ constitutes a Walrasian equilibrium. Conversely, if $(\mathbf{p},\mathbf{q})$ is a Walrasian equilibrium then $Y(\mathbf{p},\mathbf{q},\omega)=0$. To summarize, $(\mathbf{p},\mathbf{q})$ is a Walrasian equilibrium if and only if $(\mathbf{p},\mathbf{q})$ is a root of $Y(\mathbf{p},\mathbf{q},\omega)$.

Now consider the maximization problem
\begin{displaymath}
\max_{(\mathbf{p},\mathbf{q})\,\in\,\field{R}^L_+\times F(\omega)}\,Y(\mathbf{p},\mathbf{q},\omega)\,=\,\max_{\mathbf{q}\,\in\,F(\omega)}\,U(\mathbf{q})-\max_{\mathbf{p}\,\in\,\field{R}^L_+}\,V(\mathbf{p},\omega)
\end{displaymath}
If $\max_{\mathbf{q}}U(\mathbf{q})=\max_{\mathbf{p}}V(\mathbf{p},\omega)$ then $(\mathbf{p},\mathbf{q})$ is a root of $Y(\mathbf{p},\mathbf{q},\omega)$. Conversely, since $Y:\field{R}^L_+\times F(\omega)\rightarrow\field{R}_{\leq0}$, if $(\mathbf{p},\mathbf{q})$ is a root of $Y(\mathbf{p},\mathbf{q},\omega)$ then it is a maximizer of $Y(\mathbf{p},\mathbf{q},\omega)$ and
$\max_{\mathbf{q}}U(\mathbf{q})=\max_{\mathbf{p}}V(\mathbf{p},\omega)$.\hfill$\blacksquare$

\medskip
\medskip
\medskip
\medskip
\medskip

\addtolength{\baselineskip}{0.7mm}
\bibliographystyle{klunamed}
\bibliography{references}

\end{document}